\documentclass[aps,prl,preprint,showpacs,floats,byrevtex]{revtex4}
\usepackage[T1]{fontenc}
\usepackage[latin1]{inputenc}
\usepackage{graphicx}
\usepackage{amssymb}

\makeatletter

\usepackage{amsmath,amssymb,graphicx}
\def\det{\mathop{\operator@font det}\nolimits}
\def\Re{\mathop{\operator@font Re}\nolimits}
\def\Im{\mathop{\operator@font Im}\nolimits}

\makeatother
\begin{document}

\title{Electric response of DNA hairpins to magnetic fields}

\author{Juyeon~Yi}

\email{jyi@pusan.ac.kr}

\affiliation{Department of Physics, Pusan National University, Busan 609-735,
Korea}

\author{Henri Orland}

\email{henri.orland@cea.fr}

\affiliation{Service de Physique Th\'{e}orique, CEA-Saclay, 91191 Gif-sur-Yvette
Cedex, France}

\date{September 15, 2004}

\begin{abstract}
We study the electric properties of single-stranded DNA molecules with
hairpin-like shapes in the presence of a magnetic flux. It is
shown that the current amplitude can be modulated by the applied
field. The details of the electric response strongly depend on the
twist angles. The current exhibits periodicity 
for geometries where the flux through the plaquettes
of the ladder can be cancelled pairwise (\textit{commensurate}
twist). Further twisting the
geometry and changing its length causes complex aperiodic
oscillations. We also study persistent currents: They reduce to
simple harmonic oscillations if the system is commensurate,
otherwise deviations occur due to the existence of closed paths
leading to a washboard shape.
\end{abstract}

\maketitle
\textit{Introduction.}--- Recently, conduction properties of molecules
have been intensely investigated. This has been fueled by a strong motivation
to replace semiconductors with molecules that can save manufacturing
expenditure due to their self-assembly properties. One of the significant
steps towards molecule-based electronic devices has been made through
DNA conduction measurements. Superconductivity was observed, but was
supposed to be due to the proximity effect of the superconducting
leads~\cite{kasumov}. A random base sequenced DNA was shown to behave
as an insulator~\cite{pablo}, whereas the homogeneous poly(G)-poly(C)
molecule have a large gap in their current-voltage~(IV) characteristics~\cite{porath}.
Also, Ohmic behavior of DNA ropes was studied by Fink \textit{et
al.}~\cite{fink}.

On the other hand, in small systems where the phase coherence of electrons
can be maintained, quantum interferences play a key role in determining
the characteristic properties. Indeed, a large number of devices that have been proposed,
such as switches and transistors, are based on
the wave nature of electrons~\cite{alamo}. One way to observe the
interferences in a controllable way is to apply a magnetic field.
It comes into play when the systems can accommodate non-simply-connected
paths; the presence of the flux tube gives a phase shift in the wave
packet of the particle and changes the interference pattern\cite{AB}.

Despite abundant literature on DNA conduction, less attention has
been paid to the existence of interference effects in this system.
Molecules however have a huge variety in their structures allowing
for different interference events. For instance, in a double
stranded DNA, base pairs wind about the helical axis, which can be
modelled as electrons travelling on a twisted ladder-like
structure. When a magnetic flux is present, there exist
trajectories enclosing a finite flux, which may affect the
physical properties of the system. Furthermore, if one considers
single stranded DNA or RNA, one may
observe even more interesting geometries with loops and bulges. %
\begin{figure}[b]
\centerline{\includegraphics[width=0.75\columnwidth]
{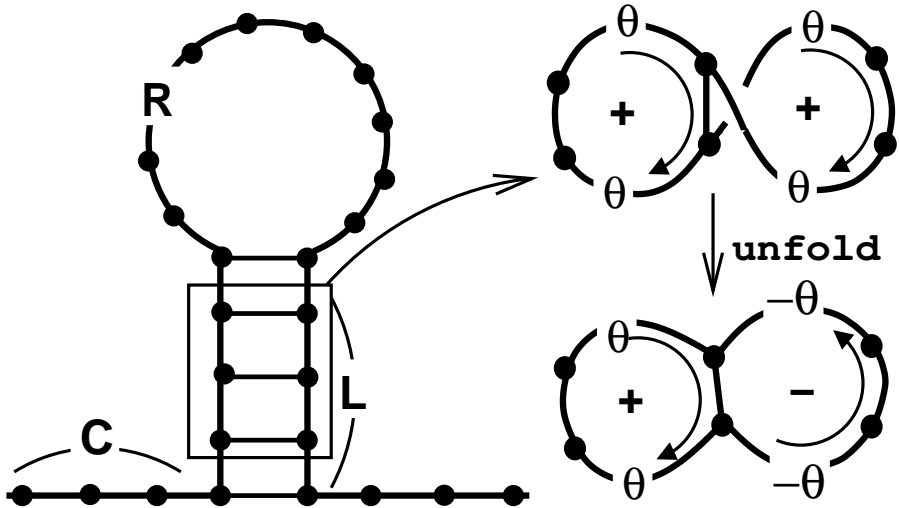}}
\caption{Schematic picture of a hairpin-structured single-stranded DNA where bases are
paired in the L(adder) sector, unpaired in
the R(ing) and in the side C(hains). The right figures are
shown to illustrate effective-flux configurations in the ladder with
the twisted rung-bonds.}
\label{fig:figure1}
\end{figure}
 The purpose of this paper is to elucidate the effects of interference
on the electronic properties of these molecules. As a minimal
model, we consider a hairpin-like structure in the presence of a
magnetic flux, as depicted in Fig.~1. The bases are coupled to
each other via electron hopping and Coulomb repulsion. For
simplicity, we consider them as identical except for the hopping
constants. We evaluate the currents through the structures when
they are coupled to electron reservoirs. The magnetic flux applied
to the system is shown to modify the current amplitude. It is also
shown that the characteristics depend on the twist angles. For a
geometry where the fluxes through the plaquettes of the ladder are
cancelled pairwise~(we refer to it as \textit{commensurate}
ladder), the current oscillates as a function of the flux through
the loop in units of flux quantum $f$, with period $f=1$, as in a
single-loop~\cite{openring}. Twisting the geometry or changing its
length changes the periodic oscillation into complex aperiodic
ones. We have also evaluated the persistent currents~(PC) of the
system~\cite{pc} which with a commensurate ladder behaves as that
of a single loop. For an \textit{incommensurate} ladder,
non-vanishing contributions are built-up by the closed paths
embracing the ladder, leading to washboard-shaped PC.

\textit{System.}--- We assume that the system is described by a Hubbard-type
model for spinless fermions~\cite{yi}: \begin{equation}
{\mathcal{H}}=\sum_{\langle i,j\rangle}[t_{ij}c_{i}^{\dagger}c_{j}+U_{ij}n_{i}n_{j}+\mbox{h.c}],\end{equation}
where $n_{i}=c_{i}^{\dagger}c_{i}$ with $c_{i}$ the operator annihilating
the Fermionic particle at site $i$, and the sum is over the nearest
neighboring pairs. This effective Hamiltonian was used to explore
the conductance gap observed in the experiments of ref.\cite{porath}
and the gapless states in the engineered DNA~\cite{mdna}. Partitioning
the hairpin into C(hain), L(adder), and R(ing) sectors~(see Fig.~1),
we assign the hopping parameters along the contour as follows: \[
t_{ij}=\left\{ \begin{array}{ccc}
t_{\parallel}\hspace{2.2cm}\mbox{for}\hspace{0.2cm}i-j=\pm1,(i,j)\in C\\
t_{\parallel}\exp(\pm2\pi if)\hspace{0.3cm}\mbox{for}\hspace{0.2cm}i-j=\pm1,(i,j)\in R\\
t_{\parallel}\exp(\pm2\pi i\theta_{i})\hspace{0.2cm}\mbox{for}\hspace{0.2cm}i-j=\pm1,(i,j)\in L\end{array}\right.\]
 In addition, the ring hoppings in the ladder are characterized by
\begin{equation}
t_{i<j}=t_{\perp}\exp(2\pi if)\label{runghop}\end{equation}
 and $t_{i>j}=t_{i<j}^{*}$ with $i$ and $j$ satisfying $i+j=N_{C}+N_{L}+N_{R}+1$
with $N_{C,L,R}$ the number of sites in the chain, the ladder, and
the ring, respectively.

Here the phases accompanying the hopping integral are due to the presence
of the threading flux $f$ and are defined as $t_{ij}=t_{\parallel}e^{i{\mathcal{A}}_{ij}}$
with ${\mathcal{A}}_{ij}=(2\pi/\Phi_{0})\int_{i}^{j}{\textbf{A}}\cdot d\vec{\ell}$.
For our system, the phase factor in the ring takes a simple form as
${\mathcal{A}}_{ij}=2\pi\Phi/N_{R}\Phi_{0}\equiv2\pi f/N_{R}$ with
$\Phi$ measuring the total flux through the ring in units of the
flux quantum $\Phi_{0}$. The magnetic flux penetrating the unit cell
in the ladder introduces the phase factor $\theta_{i}$ given by
\begin{equation}
\theta_{i}=S_{i}Ba^{2}/\Phi_{0}
\end{equation}
where $a$ is the effective path length between bases,
$S_{i}$ is the factor accounting for the area variation along the sites due to the
twist of base: \begin{equation}
S_{i}=\frac{1}{2}\left[\cos\left(\frac{2\pi i}{{\mathcal{T}}}\right)+\cos\left(\frac{2\pi(i+1)}{{\mathcal{T}}}\right)\right],\end{equation}
and $2\pi/{\mathcal{T}}$ is the twist angle that can be varied
from sample to sample, for example, by changing the amount of added
gyrase during preparation. Relating the magnetic fluxes through the
loop and the unit cell of the (flattened) ladder through the area
ratio $\gamma=4\pi/(N_{R}+2)^{2}$,
we define $\theta_{i}=\gamma S_{i}f$.

Let us consider the case of strong Coulomb repulsion where the ground
state of the system at half-filling contains charge density waves
$\langle n_{i}\rangle=(1+\Delta\cos(\pi i+\psi))/2$. At low
temperature ($U/k_{B}T\gg 1$), one can safely consider the
mean-field version of the Hamiltonian \begin{equation}
{\mathcal{H}}^{MF}=\sum_{\langle i,j\rangle}[t_{ij}c_{i}^{\dagger}c_{j}+\mbox{h.c}]+{\widetilde{U}}\sum_{i}\cos(\pi i+\psi)n_{i},\end{equation}
 where the renormalized strength of the repulsion is given by ${\widetilde{U}}=U\Delta$.
Therein, we have made the asumption that the repulsion between bases
in pair in the ladder is negligible and $\psi$ is constant over
the system. If this is not the case, an intricate point arises; specifically,
when $N_{R}$ is odd, the density waves must be distorted and have
abrupt spatial changes in $\psi$, giving rise to a kink (or a soliton).
The existence of kinks and their dynamics in an odd-numbered ring
can be another problem of interest. However,for the sake of simplicity,
we leave it for future studies.

\textit{Transmission and currents}---Let us now study the physical
properties of the system by evaluating its electric conductance. To
this end, we consider electrodes coupled to the system as in Fig.~1,
introducing the self-energy correction $\Sigma_{\mbox{lead}}$ on
the edges of the chain as \begin{equation}
{\mathcal{H}}_{c}=\Sigma_{\mbox{lead}}(c_{1}^{\dagger}c_{1}+c_{N}^{\dagger}c_{N}).\end{equation}
 There is in fact some energy dependence in $\Sigma_{\mbox{lead}}$
but considering the bulky electrodes mostly used in experiments, we
regard it as energy independent and identical for both electrodes.
Evaluation of the Green function of the system, $G(E)=(E-{\mathcal{H}}^{MF}-{\mathcal{H}}_{c}-I0^{+})^{-1}$,
leads to the transmission coefficient $t$ from the left to the right
lead as~\cite{fisherlee}\begin{equation}
t(E)={\mbox{T}r}\Gamma_{\ell}G(E)\Gamma_{r}\label{FL}\end{equation}
 where the coupling matrices are given by $\Gamma_{\ell(r)}=\mbox{Im}(\Sigma_{\mbox{lead}})\delta_{\ell(r),1(N)}$.
The corresponding currents driven by the finite voltage bias are
obtained by the Landau-B\"{u}ttiker formula~\cite{landauer}\begin{equation}
I=\int_{-\infty}^{\infty}dE[f(E-\mu_{\ell})-f(E-\mu_{r})]|t(E)|^{2},\label{LB}\end{equation}
 where $f(E)$ is the Fermi-Dirac distribution function, $\mu_{\ell}$
and $\mu_{r}$ are the chemical potential of the left and right leads
respectively, the difference of which is controlled by the applied
bias voltage $V$ as $\mu_{\ell}-\mu_{r}=eV$. %
\begin{figure}[b]
\includegraphics[width=0.75\columnwidth]{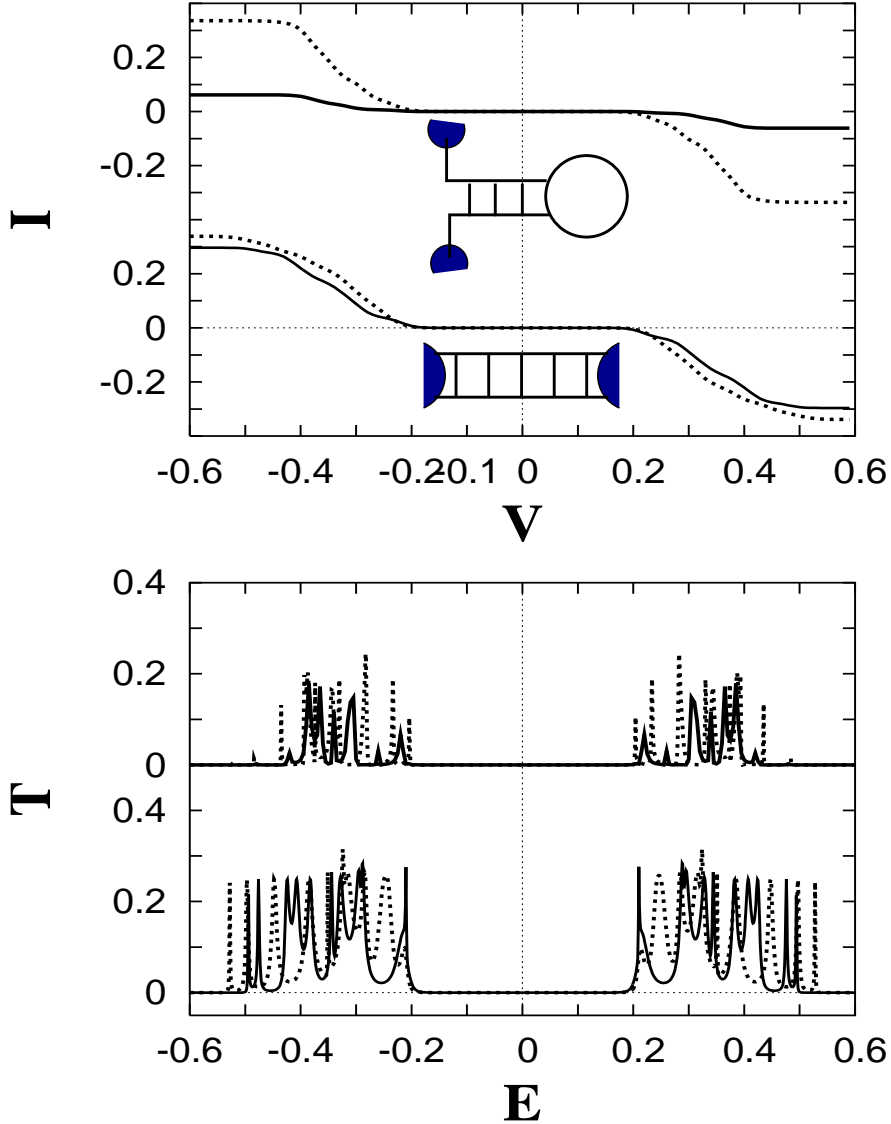}
\caption{The transmission vs energy (lower panel) and IV curves (upper panel)
for the ladder and the hairpin (see the schematics in the upper panel).
We take $30$ base pairs for the ladder, and $N_{C}=4,N_{L}=14,N_{R}=8$
for the hairpin, and ${\mathcal{T}}=10$. The band parameters used
throughout this paper are $t_{\parallel}=0.2$ eV, $t_{\perp}=0.1$
eV, ${\widetilde{U}}=0.2$ eV . The dotted curves are for $f=0$ and
the solid curves for $f=0.5$.}
\label{fig:figure2}
\end{figure}

Figure 2 displays transmissions and currents for the two configurations
(see the upper panel), the ladder and the hairpin with their ends
coupled to external leads. The transmission windows for the hairpin
gets narrower than that for the ladder, due to the smaller number
of bonds. For the ladder, when magnetic fields are absent, the IV
curve agrees with the IV characteristics for the homogeneous poly(G)-poly(C)
molecules with a large gap~\cite{porath}. When a magnetic field
with a strength $f=0.5$ is applied, the current is suppressed. This
is also the case for the hairpin, with a different degree of suppression.

\begin{figure}[t]
\centerline{\includegraphics[width=0.95\columnwidth]
{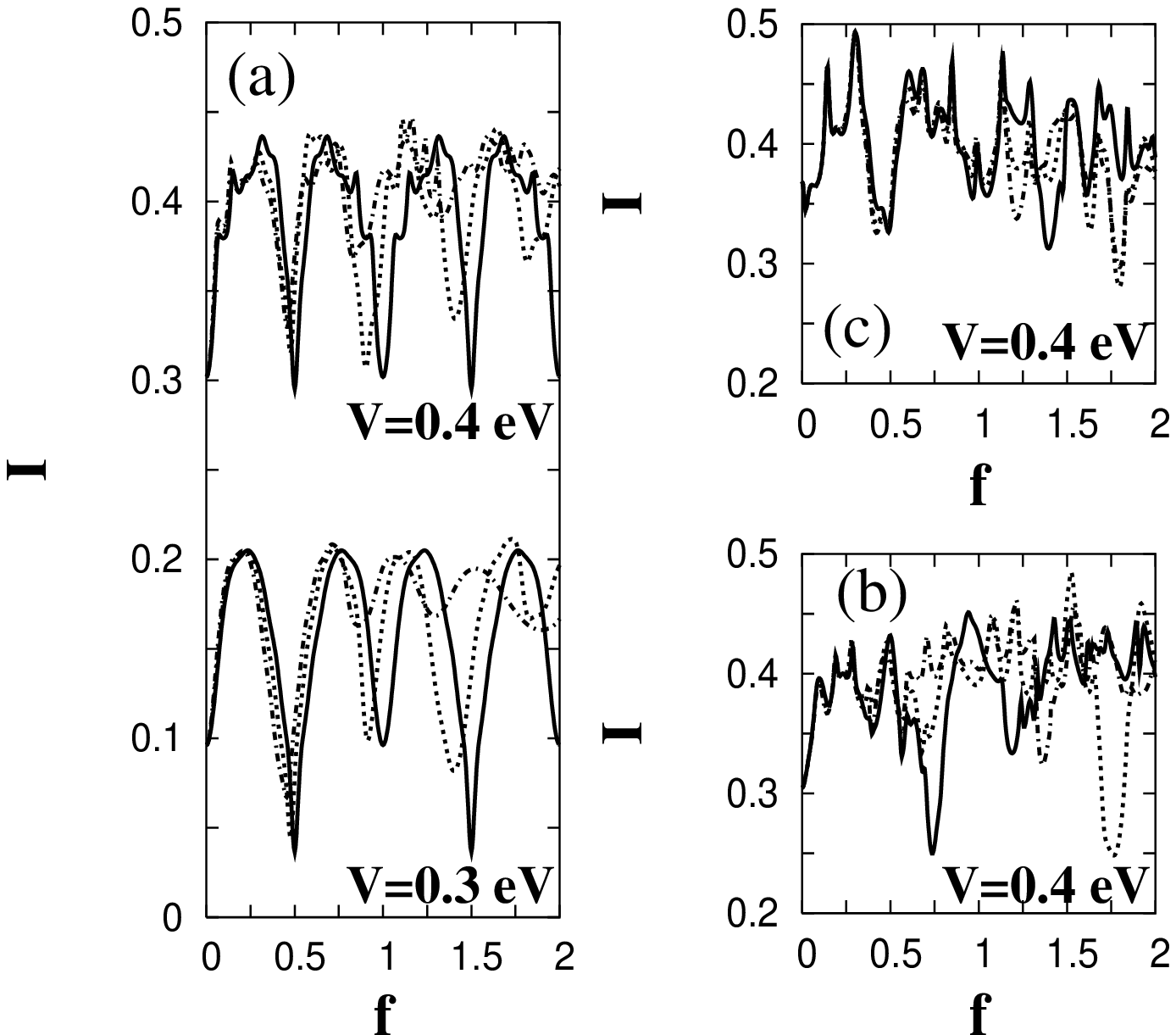}}
\caption{The current vs magnetic flux at a given voltage bias for hairpins
in different configurations with $N_{c}=4$: (a) $N_{L}=14$ and $N_{R}=8$;
(b) $N_{L}=30$ and $N_{R}=14$; (c) $N_{L}=30$ and $N_{R}=26$ with
${\mathcal{T}}=10$ (solid lines), ${\mathcal{T}}=10.5$ (
dotted lines), and ${\mathcal{T}}=10.5$ (dot-dashed lines).}
\end{figure}

To see how the magnetic field alters the current amplitude, let us
fix the voltage bias, for example, $V=0.3$ eV and $V=0.4$ eV for
the hairpin configuration of Fig.~2. Figure~3 (a) demonstrates how
the corresponding currents oscillate with varying flux. Here we present
three curves for a given voltage, each one for a different twist angle.
The interesting feature is that for ${\mathcal{T}}=10$~(solid lines)
the oscillation is periodic with period $f=1$. On the other hand,
oscillations are not complete even until $f=2$ for the case ${\mathcal{T}}=10.5$~(dotted
lines) and ${\mathcal{T}}=11$~(dot-dashed lines). One can also notice
that for ${\mathcal{T}}=10$, the current dips are located at the multiples
of $f=0.5$. Taking other twist angles moves the dips. This twist-angle
dependence can be understood as follows: Consider the area factor
$S_{i}$. The flux through a closed path $c_{j}$ in the ladder is
given by $\sum_{i\in c_{j}}S_{i}\equiv=S^{(j)}$. For the hairpin
configuration~($N_{C}=4,N_{L}=16,{\mathcal{T}}=10$), every path
$c_{k}$ has its respective counter path $c_{j}$ in such
a way that pairwise cancellation of the flux occurs since $S^{(k)}=-S^{(j)}$.
Consequently, the flux through the ladder gives null effects. In that
case, in terms of the magnetic flux, the geometry is topologically
equivalent to that of a single loop connected to the electron reservoirs
(see Ref.~\cite{openring}): The transmission of this configuration
has been shown to oscillate with period $f=1$. When twisting the structure
slightly out of the \textit{commensurate} configuration ~(for example,
${\mathcal{T}}=10.5$), such a perfect cancellation does not occur,
leading to complex aperiodic oscillations (within the flux interval).
One can also expect that adding more plaquettes must yield similar
features. In Fig.~3 (b) and (c), doubling the size of the ladder,
we plot the currents at $V=0.4$ eV. Neither of the twist angles leads
to periodicity. Also, the current enhancement by finite flux $0<f<0.5$
can be noticed. Let us here give an order of magnitude of the strength
of the magnetic field for the effects to be measurable. For small
loops with radius $50 \AA$ (corresponding approximately to 8 bases),
in order to achieve $f=0.5$, one needs fields in the
range of thousands of Tesla. Considering that at present the maximum
field strength is about 100 Tesla ~\cite{magfield}, it would be
easier to realize a larger loop. In Fig.~3 (c), we take 24 bases in
the loop, making the area about ten times bigger than that in Fig.~3(a).
Even if the field strength can be in the realizable range, the effects
remain still significant and are not washed out by the enlarged loop-size.
\begin{figure}[t]
\centerline{\includegraphics[width=0.85\columnwidth]
{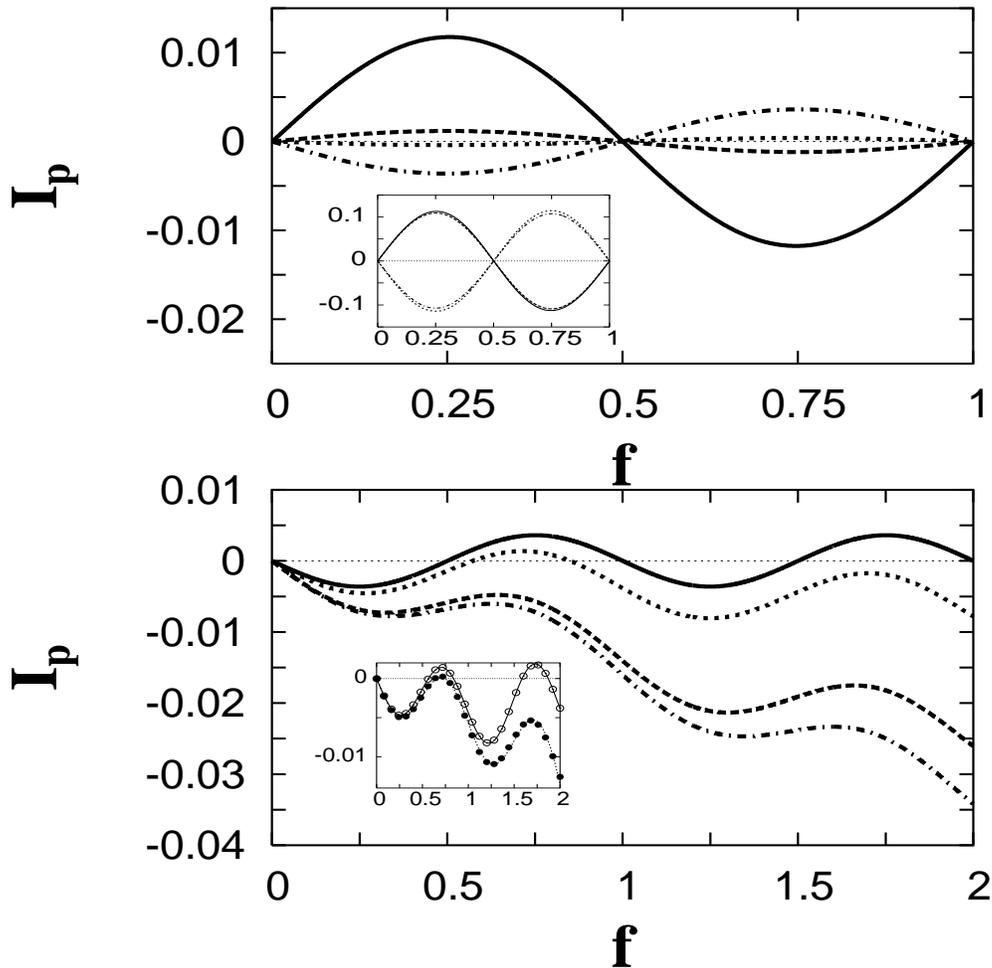}}
\caption{The persistent currents vs flux: The upper panel shows the PC of
the hairpins ($N_{C}=4,N_{L}=14$) with different number of bases
in the loop, $N_{R}=6$~(solid line), $N_{R}=8$~(dashed line), $N_{R}=10$~(
dot-dashed line), and $N_{R}=12$~(dotted). While the oscillation
is periodic in $f$, the amplitude decreases as the loop size increases.
In the inset, we show that
the curves coalesce when using a scaling $I_{p}e^{0.56N_{R}}$.
The lower panel displays
simple harmonic oscillations of PC for $N_{C}=4$, $N_{R}=6$ and $N_{L}=14$
(solid line), obtained by changing the ladder size $N_{L}=18,22,26$~(
dotted, dashed, dot-dashed, respectively). The inset shows
the fits according to Eq.~(\ref{approxpc}) for $N_{L}=2$ and
$N_{L}=4$.}
\end{figure}

\textit{Persistent currents}--- Now we evaluate the persistent currents
when the system is isolated. This allows to assess how much the magnetic
fields alter the energy levels (whereas the magnetic-flux effect demonstrated
above mixes the influence on wavefunctions and on the energy spectrum).
This becomes clear when writing the Green function in an
expansion of eigenfunctions: \begin{equation}
G(x',x;E)=\sum_{n}\frac{\psi_{n}^{*}(x')\psi_{n}(x)}{E-E_{n}-\Sigma(E)},
\end{equation}
leading to the transmission coefficient given in Eq.~(\ref{FL}):
Here the magnetic flux affects not only the energy spectrum $E_{n}$
but also the wavefunction $\psi(x)$. Let us take a simple example
for understanding persistent currents. Consider a ring of radius $R$
threaded by a flux $\phi$ (in units of the flux quantum $\Phi_{0}$).
The angular momentum of free electrons traveling on the ring is shifted
by the magnetic flux as $\langle n|L|n\rangle=\hbar(n+\phi)$ with
$\langle n|{\mathcal{A}}|n\rangle$ denoting the expectation value
of an operator ${\mathcal{A}}$ in the $n$th eigenstate. The energy eigenvalues
are then given by ${\mathcal{E}}_{n}=\hbar^{2}(n+\phi)^{2}/2mR^{2}$.
At zero-temperature, the evaluation of currents in the ground state
is straightforward \begin{equation}
I_{p}=-\sum_{n}\frac{e}{2\pi R}\langle n|
{\widehat{v}}|n\rangle=-\frac{e}{2\pi\hbar}
\sum_{n}\frac{\partial{\mathcal{E}}_{n}}{\partial\phi},\label{pc}\end{equation}
where ${\widehat{v}}$ is the velocity operator, and the sum is over
all the occupied levels, $n=0,\pm1,\dots,\pm(N_{e}/2-1),-N_{e}/2$
for an even number $N_{e}$ of electrons and $n=0,\pm1,\dots,\pm(N_{e}-1)/2$
for odd $N_{e}$ . It is clear that the ground state carries
currents so that if dissipation is absent in the system, the currents
keep flowing, and are thus named persistent currents. The currents
oscillate with the magnetic flux with a period of $\phi=1$, and
their amplitudes decrease as $1/R$. When band gaps are present in
the system (due for example to Coulomb repulsion or coupling to
lattice distorsion), the persistent currents are suppressed exponentially
with increasing $R$\cite{loss}. In this case, one has \begin{equation}
I_{p}=(-1)^{N_{e}}I_{0}e^{-\alpha R}\sin(2\pi f)\label{pcgap}\end{equation}
 with $\alpha=0$ for vanishing gap. Here the sign of the current depends
on the parity of $N_{e}$\cite{pc,loss}.

Our system (a loop coupled to a ladder) has a number of closed paths
or rings, with different areas and sizes. There are contributions of
persistent currents from all the rings, and therefore, $I_{p}=\sum_{i}I_{i}$,
where $I_{i}$ denotes the current flowing in the $i$-th ring of
size $R_{i}$, enclosing the flux $f_{i}$. We first evaluate the
persistent currents in the hairpin with $N_{c}=4$ and $N_{L}=14$
where the magnetic fluxes through the ladder plaquette effectively cancel
one another~(see the upper panel in Fig.~4). The currents behave
as those of a single loop; note that the sign of the current changes with
the parity of $N_{e}$~(we consider half-filled systems). Furthermore,
increasing the number of bases decreases the current amplitude. Following
the finite size scaling in Eq.~($\ref{pcgap}$), all the curves are
shown to be merged into a single one~(see the inset). Adding more
plaquettes causes a deviation from a single-component sinusoidal-curve~(see
the upper panel in Fig.~4). Since larger rings die out exponentially
, taking paths enclosing up to two unit-cells,
one obtains the approximate form of the currents
as \begin{equation}
I_{p}\approx I_{r}\sin(2\pi f)+\sum_{i}I_{s,i}\sin(2\pi\theta_{i})+\sum_{i}I_{d,i}\sin(2\pi\phi_{i}),\label{approxpc}\end{equation}
 where $\theta_{i}$ is the flux through the $i$th plaquette of the
ladder, and $\phi_{i}$ is the sum of two neighboring $\theta_{i}$'s.
For hairpins with a few plaquette in the ladder, Eq.~(\ref{approxpc})
perfectly fits the numerical evaluations (obviously, for a
hairpin with a single plaquette, the above expression is exact).

\textit{Summary and Remarks}--- We have demonstrated how a magnetic
flux could influence the electronic properties of molecules with
a twisted hairpin-like shape. The current has been shown to
oscillate with flux changes. Geometry factors such as
twist angles and the number of rungs of the ladder play a crucial
role in determining oscillation patterns. We point out that there
exists geometries where the flux through the ladder vanishes,
so that the sole cause for the oscillation is the flux
through the loop. For that case, the current of the 
ladder ressembles that of a ring. Twisting the geometry
brings about complex aperiodic oscillations. These oscillations are also 
reflected
on persistent currents. The non-vanishing contributions of the closed paths
embracing the ladder leads to washboard-shaped persistent currents.

We should note that the above mentioned results were predicted
under the assumption that electrons preserve their initial phase in their
propagation. Therefore, maintaining phase
coherence is crucial in experimental observations of these phenomena. 
The phase coherence is strongly affected
for example by temperature (or generally speaking, coupling to the
environmental degrees of freedom). For DNA molecules for which
conduction properties are measured at room-temperature, phase
coherence can still be traced. However, finite-temperature effects cause
fluctuations in the structure. Even if the effect is small,
the plaquette areas as well as the phases due to the flux become randomized.
This addresses the issue of (weak) localization that will be
considered in a future work.

\end{document}